\documentclass[runningheads]{llncs}
\usepackage[T1]{fontenc}
\usepackage{graphicx}
\usepackage{amsmath}
\usepackage{amssymb}
\usepackage{booktabs}
\usepackage{tabularx}
\begin{document}
\title{How to solve a classification problem using a cooperative tiling Multi-Agent System?}
%
%
\author{Thibault Fourez\inst{1,2}\orcidID{0000-0003-1677-6623} \and
Nicolas Verstaevel\inst{1}\orcidID{0000-0002-7879-6681} \and
Frédéric Migeon\inst{1} \and
Frédéric Schettini\inst{2} \and
Frédéric Amblard\inst{1}}
\authorrunning{T. Fourez et al.}
%
\institute{Institut de Recherche en Informatique de Toulouse, Université de Toulouse, CNRS, Toulouse INP, UT3, UT1, Toulouse, France \\
\url{https://www.irit.fr} \and
Citec Ingénieurs Conseil, Geneva, Switzerland}

\maketitle              
\begin{abstract}
Adaptive Multi-Agent Systems (AMAS) transform dynamic problems into problems of local cooperation between agents. We present \textit{smapy}, an ensemble based AMAS implementation for mobility prediction, whose agents are provided with machine learning models in addition to their cooperation rules. With a detailed methodology, we propose a framework to transform a classification problem into a cooperative tiling of the input variable space. We show that it is possible to use linear classifiers for online non-linear classification on three benchmark toy problems chosen for their different levels of linear separability, if they are integrated in a cooperative Multi-Agent structure. The results obtained show a significant improvement of the performance of linear classifiers in non-linear contexts in terms of classification accuracy and decision boundaries, thanks to the cooperative approach.

\end{abstract}
\section{Introduction}

Supervised classification problems have been extensively addressed using machine learning algorithms called classifiers. These problems are generally not linearly separable (i.e. the point clouds of the different classes are not separable by hyperplanes). Moreover, in so-called dynamic problems, new classes appear and disappear and the behavior of individuals evolves in time and space (e.g. management of smart cities and the appearance of new transport modes and behavior). Dynamic problems are often associated with ambient environments in which new devices may appear or disappear dynamically \cite{guivarch2013dynamic,perles2018amak}. Many classifiers are themselves composed of multiple linear classifiers (i.e. whose decision function is a linear combination of input variables) that they aggregate.

After positioning \textit{smapy} with respect to machine learning techniques for classifier aggregation in section \ref{sota}, we detail its operation in section \ref{smapy}. In section \ref{experiment}, we show that our MAS is able, thanks to a space exploration by four types of linear classifiers and cooperation rules, to linear toy classification problems with different levels of linear separability from the literature. \textbf{We show through this experiment that it is possible to transform a classification problem into a cooperation problem} through the exploration of the space of input variables. The results presented in section \ref{results} are discussed in section \ref{discussion}.

The main contributions of this paper are:
\begin{itemize}
    \item \textit{Smapy}, an ensemble Multi-Agent System (MAS) for online non-linear classification based on the AMAS framework
    \item An experiment on three two-dimensional problems from litterature to evaluate the performance of our MAS for different levels of linear separability
\end{itemize}

\section{Related work}\label{sota}

In this section, we situate our approach among other machine learning methods based on aggregation of classifiers. After presenting their advantages and weaknesses, we show that the AMAS approach is a constructivist and ensemble learning method. We then detail how our approach meets the objectives of \textbf{online} solving of \textbf{dynamic problems} through cooperative tiling of the input variable space.

\subsection{Aggregation of classifiers}\label{machine_learning}

Many non-linear classifiers aggregate several linear classifiers. We distinguish two types of aggregation: connectionist and ensemble approaches. The resulting algorithms can be self-constructing.

\subsubsection{Connectionism}

The connectionist approach can be illustrated by a serial electrical circuit. The input data passes successively through a chain of classifiers and the decision function of the global classifier is a composition of the decision functions of the internal classifiers. In supervised learning, the information on the output goes back up the chain of classifiers in the opposite direction.

The connectionist approach is associated with artificial neural networks (ANN) in which the information of the output data is fed back through the gradient back-propagation algorithm \cite{rumelhart1986learning}. The internal classifiers composing the neural network are linear. A network is usually composed of several layers (multi-layer perceptron). The multiplication of these layers is at the origin of deep learning.

The main limitation of the connectionist approach is the need for a large number of observations to optimize the weights associated with each internal classifier. In dynamic problems where the behavior of individuals evolves over time, new data are generally not available in sufficient number to update the learning model.

\subsubsection{Ensemble}

The ensemble approach is similar to a parallel electrical circuit. The input data passes through several classifiers in parallel, and the output of the decision function is an aggregation of the outputs of the decision functions of each classifier.

Bagging is a learning technique combining the use of bootstraps and the aggregation of prediction models. The assumption of bagging, inspired by the law of large numbers, is that averaging the predictions of several independent models reduces the variance and thus the error of the global prediction. The Random Forest algorithm \cite{breiman2001random} uses binary decision trees by adding a random draw of the input variables to be considered for each intermediate model.

Unlike bagging, boosting algorithms build a model sequentially from so-called weak models. At each step, the bad points predicted by the previous model are given a higher weight when training the current model. Adaptive Boosting (AdaBoost) \cite{schapire1995decision} uses binary decision trees with a single node and a single input variable. In Gradient Boosting \cite{friedman2002stochastic} and eXtreme Gradient Boosting (XGBoost) \cite{chen2016xgboost}, the weights of the points are no longer incremented but a cost function minimized by gradient descent allows to aggregate intermediate models to the global model. 

\subsubsection{Constructivism}

In self-constructing aggregation approach, the graph of internal classifiers (connectionist or ensemble) does not have a fixed structure. This structure is built during the learning process and can evolve over time. When changes to the classifier graph are made by the classifiers themselves, the system is self-organizing.

Algorithms such as self-constructing neural networks \cite{carmo2009constructive} lie at the intersection of connectionism and constructivism, because the structure of the layers of neurons is built incrementally.

Context-aware learning is a constructivist and almost ensemble approach. In ELLSA \cite{dato2021lifelong}, the Context agents can be seen as linear classifiers (although they can only predict one class) with cooperative rules.
\\\\
The constructivist approach allows us to design self-organizing non-linear classifiers suitable for solving dynamic problems. In this paper, we propose an ensemble context-based learning MAS. In the following paragraphs, we situate our contribution in the field of adaptative MAS.

\subsection{Multi-Agent Systems}

Multi-agent systems (MAS) are systems involving autonomous entities called agents, capable of communicating in a common environment in which each has its own perceptions and knowledge \cite{ferber1999multi}. Each agent acts locally to get closer to its objectives using its own skills.

Problem solving by MAS is the search for a balance in the interactions between agents. It is not the sum of the individual abilities of the agents, but rather the product of their interactions, which leads to the emergence of new resolution abilities. This phenomenon of emergence is regularly invoked to justify the use of MAS for the collective resolution of dynamic and complex problems.

\subsubsection{Self-organization}

Self-organization is the capacity of a system to dynamically modify its internal structure without external intervention \cite{marzo2011history}. Self-organization is said to be strong when the modifications are not the consequence of a centralized decision. In the framework of MAS, the agents can modify the global function of the system by acting on their local environment. This process of adaptation to the dynamics of the problem is similar to machine learning when the agents share a common goal.

The theory of adaptive multi-agent systems (AMAS) \cite{capera2003amas} proposes a cooperative approach to interactions between agents. The design criteria presented for these interactions guarantee a satisfactory, but not necessarily optimal, result in the resolution of the problem at hand (functional adequacy theorem).

\subsubsection{Context learning}

Context learning consists in exploring the space defined by the input variables of the model using cooperating agents. The AMAS for context learning (AMAS4CL) approach is based on the AMAS theory and more particularly on the Self-Adaptive Context Learning (SACL) \cite{boes2015self} paradigm to define the rules of cooperation between agents and proposes a structure composed of several types of agents to explore the space of the problem variables.
Algorithms based on the SACL approach are used to solve various problems such as learning by demonstration \cite{verstaevel2015principles} or Inverse Kinematics \cite{dato2021cooperative} in robotics and optimization of the operation of a heat pump \cite{boes2015self}.
SACL architectures are typically composed of three types of agents:
\begin{itemize}
    \item Context agents that define hypercubes of the input variable space. When a new point belongs to one of these zones, the corresponding Context agent is said to be activated and proposes a system decision according to its own knowledge
    \item The Percept agents which retrieve the values of the input variables (sensors) at each iteration and transmit them to the Context agents
    \item The Head agent which receives the proposals of the activated Context agents and sends them feedbacks from oracle
\end{itemize}

\begin{figure}
\includegraphics[width=\textwidth]{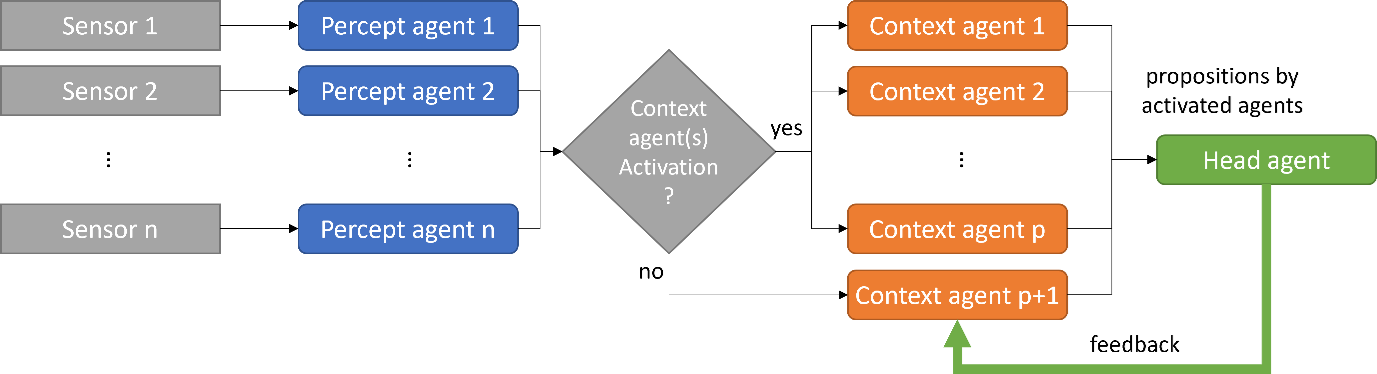}
\caption{Cooperative operation of a SACL architecture in exploration phase} \label{schema_sacl}
\end{figure}

A context learning MAS has two modes of operation: exploration (learning) which consists in tiling the input variable space by instantiating and arranging the Context agents thanks to the feedback from the Head agent, and exploitation (prediction) which consists in taking a decision without updating the system.  The functioning of this architecture is presented in Fig.~\ref{schema_sacl} for the cooperative case in the exploration phase (i.e., the optimal functioning case). When the behavior of the system is not optimal with respect to the user's objectives, the situation is said to be non-cooperative (NCS). The system must adapt to maximize the cooperation between agents to return to the cooperative state. In Context Learning, this cooperation is expressed in the sizes (i.e., the dimensions of the hypercubes), positions and knowledge of the Context agents.

The SACL approach is adapted to dynamic problems (i.e. the system adapts to changes in the distribution of the input data over time) and supports online learning. Moreover, the position and size of the agents in the space of the input variables provide additional information about the phenomenon under study and a geometric interpretation.

\section{Smapy} \label{smapy}

In our contribution, we provide each Context agent with an internal machine learning model, linear or not, with the only requirement to support online learning. Each internal model is trained on the points that have activated the corresponding Context agent and thus constitutes a local modeling (in the sense of the space of input variables) of the underlying function of the problem to solve. This context learning MAS, \textit{smapy}, has been implemented in python for the industrial needs of the research project.

\subsection{General principle}

Like other SACL type architectures, \textit{smapy} has two modes of operation: 
\begin{itemize}
    \item The exploration during which the mapping of the space of the input variables is modified according to new available labelled observations
    \item The operation during which the system uses its coverage of the space of input variables to classify a new point
\end{itemize}
In both cases, the operation of the system is iterative, and each cycle starts with a new observation. During the exploration, the activated Context agents update their internal model with the last observation after they have proposed an output class to the Head agent and received feedback (positive or not). The feedback received by a Context agent allows it to update its perception of itself within the collective through a performance metric explained in the section 3.3. It also allows him to know if he has a non-cooperative behavior with respect to the objective of the system and, if necessary, to act on itself or its neighbors to return to a cooperative state (c.f. section 3.4).

\subsection{Agents}

In this section, we present the three types of agents involved in our SACL architecture, whose relationships have been described in Fig.~\ref{schema_sacl}.

\subsubsection{Percept}

The $p$ Percept agents collect the values of the $p$ input variables of each new observation and pass them to the Context agents. They also store the observed extrema for each variable.

\subsubsection{Context}

A Context agent $l$ defines a hypercube in the $p$-dimensional space of input variables. For each dimension $j$, it has two parameters $r_{l,j,0}^t$ and $r_{l,j,1}^t$ that define the lower and upper bounds of an activation interval at iteration $t$. The agent can compute at any time $v_l(T)$, the volume of its activation hypercube at iteration $T$, according to the following formula: 
\begin{equation}
    v_l(T)=\prod_{j=1}^p(r_{l,j,1}^t-r_{l,j,0}^t)
\end{equation}
The Context  agent also has a confidence level $c_l(T)$ at iteration $T$, depending on its history $\mathcal{H}_l^T$ (set of its activation cycles since its creation), its class proposals $\hat{y}_l^t$t for observations $y^t$ on this history, and two external parameters $F_+$ and $F_-$ that respectively weight the positive and negative feedbacks of the agent Head:
\begin{equation}
    c_l(T)=\sum_{t\in\mathcal{H}_l^T}(F_+*\mathbb{1}_{\hat{y}_l^t=y^t}-F_-*\mathbb{1}_{\hat{y}_l^t\ne y^t})
\end{equation}
From its two terms, we define the score $s_l(T)$ of a Context agent at iteration $T$ using a normalization function $N_c$ which is an external parameter of \textit{smapy}:
\begin{equation}
    s_l(T)=N_c\circ c_l(T)
\end{equation}
Finally, the Context agent has an internal classification model learned from the observations that activated it. The python implementation of \textit{smapy} makes it possible to use models in the \textit{scikit-learn} fashion if they support online learning to adapt the agent to new observations. For the rest of this paper, we define several properties of Context agents:
\begin{definition}\textbf{\emph{(Expansion/retraction)}}
    A Context agent expands (resp. retracts) by a factor $\alpha$ when it increases (resp. decreases) its boundaries to multiply its volume by $1+\alpha$ (resp. $1-\alpha$).
\end{definition}
\begin{definition}\textbf{\emph{(Push)}}
    A Context agent $l_1$ pushes a Context agent $l_2$ when $l_2$ retracts so that the previous intersection of $l_1$ and $l_2$ is completely outside $l_2$ (and thus contained only within $l_1$).
\end{definition}
\begin{definition}\textbf{\emph{(Absorption)}}
    A Context agent $l_1$ absorbs a Context agent $l_2$ when $l_1$ expands to completely contain the area covered by $l_2$ and the agent $l_2$ is destroyed.
\end{definition}
\begin{definition}\textbf{\emph{(Point exclusion)}}
    A Context agent $l_1$ excludes an observation $y$ when $l_1$ retracts so that $y$ ends up outside $l_1$. Point exclusion is controlled by an external Boolean parameter $E$.
\end{definition}
\begin{definition}\textbf{\emph{(Overlapping index)}}
    The overlapping index $o_{l_1,l_2}$ is the ratio of the volume of the intersection of two Context agents $l_1$ and $l_2$ to the minimum of the volumes of these agents:
    \begin{equation}
        o_{l_1,l_2}=o_{l_2,l_1}=\frac{v_{l_1 \cap l_2}}{\min(v_{l_1}, v_{l_2})}
    \end{equation}
\end{definition}

\subsubsection{Head}

The Head agent supervises the cooperation of the Context agents. At each iteration, it selects the class proposed by the activated Context agent with the highest score (and proceeds by vote in case of a tie) and sends feedbacks to all the agents activated during the exploration phase (c.f. section 3.3). The Head agent can also create new Context agents in case of system incompetence (c.f. section 3.4).

\subsection{Feedback}

When the Context agents are activated, they propose a prediction to the Head agent. The latter selects the prediction of the agent with the highest score. During the exploration phase (learning), the Head agent sends feedbacks to the Context agents which have proposed a prediction:
\begin{itemize}
    \item If the prediction is good (with respect to the label of the new point), then the confidence of the context agent increases by $F_+$ and it expands by a factor $\alpha$ (external parameter)
    \item If the prediction is bad, then the confidence of the context agent decreases by $F_-$.  If point exclusion is allowed (i.e., $E$ is true), then the context agent excludes the new point. Otherwise, the Context agent's local model is fine-tuned with the new point (in the sense of online learning), and it retracts by a factor $\alpha$
\end{itemize}

\subsection{Non-cooperative situations}

\begin{figure}
\centering
\includegraphics[width=0.7\textwidth]{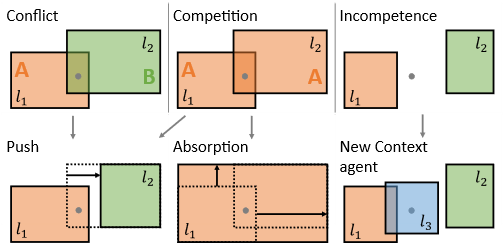}
\caption{Schematic of NCS (top row) and their resolution (bottom row) for Context agents $l_1$, $l_2$ and $l_3$ predicting classes $A$ and $B$} \label{schema_ncs}
\end{figure}

The objective of AMAS is to transform the initial problem into a problem of cooperation between agents. Non-cooperative situations (NCS) are states during which the behavior of the system must evolve to reach the goal set by the user. In context learning, this results in the rearrangement of the Context agents to improve the tiling of the input variable space. In this section, we present and schematize in Fig.~\ref{schema_ncs} the three types of NCS that can occur during the operation of \textit{smapy} and their resolution.

\subsubsection{Incompetence}

Incompetence occurs when no Context agent has been activated:
\begin{itemize}
    \item Exploration: a new Context agent is created around the new point and any NCS generated are resolved. The initial radius of the new agent is controlled by an external parameter $R$
    \item Exploitation: the closest Context agent to the new point (in the sense of the Euclidean distance between the point and the agent's boundary) proposes its prediction
\end{itemize}
	
\subsubsection{Competition}

Competition occurs during the exploration phase when two activated Context agents propose the same prediction (in this case, the same class):
\begin{itemize}
	\item If an overlapping threshold is defined through the external parameter $O$, and if the overlapping index of the two agents is greater than this threshold, the agent with the higher score absorbs the other
	\item Otherwise, the Context agent with the higher score pushes the other agent
\end{itemize}

\subsubsection{Conflict}

Conflict occurs during the exploration phase when two activated Context agents propose different predictions. The agent with the higher score then pushes the other agent.

\section{Comparison of linear classifiers alone with context learning approach}

In this section, we present the experiment of comparing four types of linear classifiers and instances of \textit{smapy} with these same models inside Context agents. The linear classifiers used in this paper are : 
\begin{itemize}
    \item Logistic regression (ridge \cite{hoerl1970ridge}, LASSO \cite{tibshirani1996regression} or ElasticNet \cite{zou2005regularization})
    \item Linear Support Vector Machine (SVM) \cite{cristianini2000introduction}
    \item Passive Agressive algorithms \cite{crammer2006online} (PA-I and PA-II)
\end{itemize}
The motivation of this experiment is to verify if the transformation of a classification problem into a multi-agent cooperation problem allows to improve the performances of the latter.

\subsection{Input data}

The experiment is conducted on three two-dimensional binary classification toy datasets included in the scikit-learn library \cite{scikit-learn} and regularly used for model comparison purposes:
\begin{itemize}
    \item Moons: Two interleaved point clouds explained by the two variables (\texttt{noise=0.3})
    \item Circles : A circular point cloud embedded in another ring-shaped cloud (\texttt{noise=0.2}, \texttt{factor=0.5})
    \item Linearly separable : Two point clouds with a linear boundary explained by only one of the two variables
\end{itemize}
Each data set contains 100 points. They are centered and reduced using the scikit-learn \texttt{StandardScaler} model before learning.

\subsection{Experimental protocol}\label{experiment}

\begin{table}
\centering
\setlength\tabcolsep{4pt}
\begin{minipage}{0.48\textwidth}
\centering
\caption{List of value grids for the search of optimal combinations of parameters of the studied linear models (scikit-learn implementation)}
\label{grid_linear_model} 
\begin{tabular}{llll}
\toprule
\textbf{Parameter} & \multicolumn{3}{c}{\textbf{Grid of values}} \\
\midrule
\multicolumn{4}{c}{\textsc{Logit \& Linear SVM}}                  \\
\texttt{alpha}          & 0.0001      & 0.001      & 0.01           \\
\texttt{penalty}        & $l_1$       & $l_2$      & ElasticNet    \\
\midrule
\multicolumn{4}{c}{\textsc{PA-I \& PA-II}}                                   \\
\texttt{C}              & 0.5         & 1.0        & 2.0            \\
\bottomrule
\end{tabular}
\end{minipage}%
\hfill
\begin{minipage}{0.48\textwidth}
\centering
\caption{List of value grids for finding the optimal combinations \textit{smapy} parameters} 
\label{grid_smapy} 
\begin{tabular}{llll}
\toprule
\textbf{Parameter} & \multicolumn{3}{c}{\textbf{Grid of values}} \\ 
\midrule
$R$                & 0.1            &  0.2           & 0.5          \\
$O$                & 0.2            & 0.5            &              \\
$E$                & False          & True           &              \\
$N_c$              & Sigmoid        &                &              \\
$\alpha$           & 0.0            & 0.1            & 0.2          \\
$F_+$              & 1.0            &                &              \\
$F_-$              & 0.5            & 1.0            & 2.0          \\
\bottomrule
\end{tabular}
\end{minipage}
\end{table}

\subsubsection{Step 1}

First, we search for the optimal combination of parameters for each of the four linear classifiers among a grid of parameters presented in the table \ref{grid_linear_model} using a five-fold cross validation.

\subsubsection{Step 2}

Once these combinations are obtained, we search for the optimal combination of \textit{smapy} parameters for each linear classifier among a grid of parameters presented in the table \ref{grid_smapy} with a five-fold cross validation. The Context agents of the \textit{smapy} instances have as internal model the corresponding linear model, trained with the parameters of its previously obtained optimal combination.
\\\\
This protocol is repeated for each dataset to obtain 12 \textit{smapy} instances and 12 corresponding linear classifier instances, all of which have been optimized by cross-validation. The optimized \textit{smapy} instances are then compared with the linear classifiers using two evaluation metrics:
\begin{itemize}
    \item Classification accuracy (multi-class) averaged over the five iterations of cross-validation (step 1 for linear models, step 2 for \textit{smapy})
    \item Decision boundaries of the models (linear or \textit{smapy}) trained with the best parameter combinations obtained by cross-validation
\end{itemize}

\section{Results}\label{results}

In this section, we present comparative results between linear classifiers alone and \textit{smapy} instances according to the two metrics introduced previously (section \ref{experiment}).

\subsection{Classification accuracy}

\begin{table}
\caption{\label{accuracies}Comparison of the classification accuracies obtained for each classifier alone or inside \textit{smapy}}
\centering
\begin{tabularx}{\textwidth}{l*{3}{>{\centering \arraybackslash}XX}}
\toprule
           & \multicolumn{2}{>{\centering \arraybackslash}l}{\textbf{Moons}} & \multicolumn{2}{>{\centering \arraybackslash}l}{\textbf{Circles}} & \multicolumn{2}{>{\centering \arraybackslash}l}{\textbf{Linearly Separable}} \\ \midrule
           & Alone    & MAS             & Alone     & MAS              & Alone            & MAS             \\ \midrule
Logit      & 0.83     & \textbf{0.89}   & 0.49      & \textbf{0.83}    & \textbf{0.92}    & 0.91            \\
Linear SVM & 0.86     & \textbf{0.89}   & 0.53      & \textbf{0.83}    & 0.90             & \textbf{0.91}   \\
PA-I       & 0.82     & \textbf{0.89}   & 0.53      & \textbf{0.83}    & 0.89             & \textbf{0.90}   \\
PA-II      & 0.84     & \textbf{0.87}   & 0.53      & \textbf{0.83}    & \textbf{0.89}    & \textbf{0.89}   \\ \bottomrule
\end{tabularx}
\end{table}

Table \ref{accuracies} shows no significant difference in accuracy after switching to MAS with the dataset \textit{Linearly separable}. The linear classifiers already achieve a high score despite the noise in the data.

For the other two datasets, we observe an improvement of the accuracy with the \textit{smapy} instances. In particular, we observe a $\sim30\%$ improvement for the dataset \textit{Circles}, for which the linear classifiers alone give a result close to chance (50\%). The poor performance of these classifiers is explained by the low linear separability of the dataset. However, we see that the integration of these classifiers in an SMA allows to approach the quadratic boundary of the dataset.

The accuracies obtained on the dataset \textit{Moons} (intermediate case in terms of linear separability) are slightly better with the MAS approach, but the linear classifiers alone allow to obtain satisfactory scores.

\subsection{Decision boundaries}

\begin{figure}
\includegraphics[width=\textwidth]{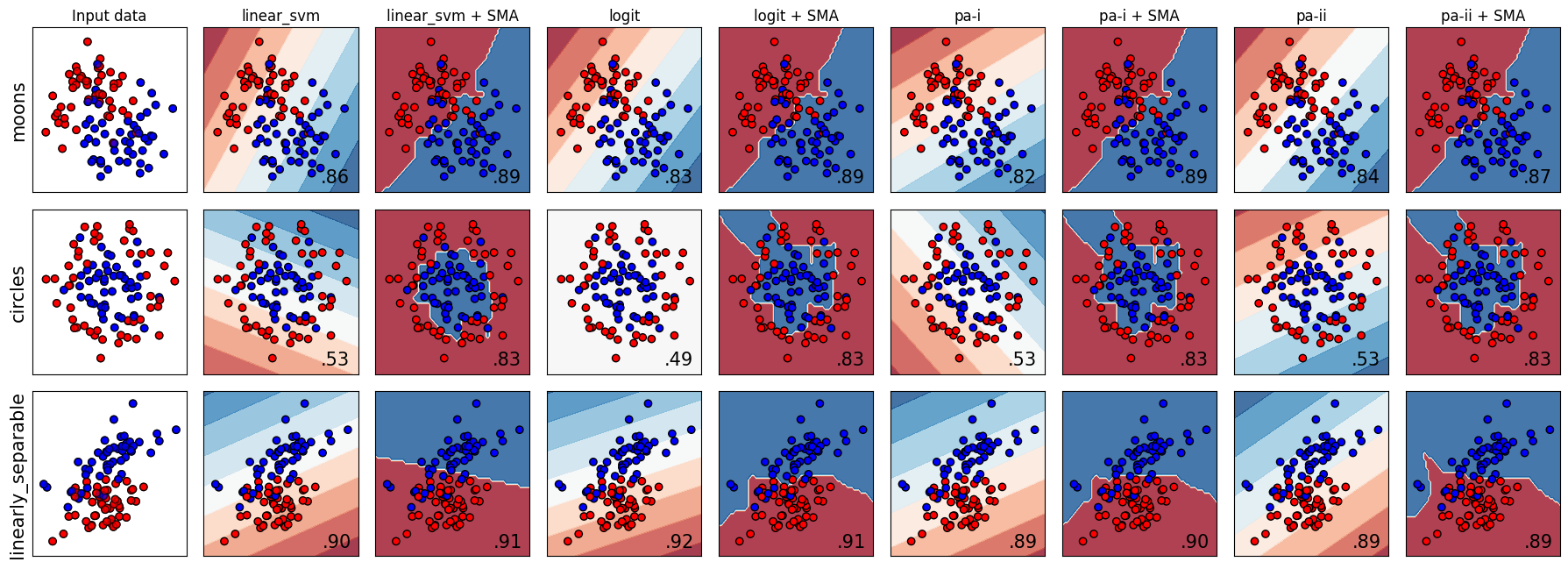}
\caption{Decision boundaries and classification accuracy obtained for each dataset (rows) and for each linear classifier alone or in a \textit{smapy} instance (columns)} 
\label{decision_boundaries}
\end{figure}

Fig.~\ref{decision_boundaries} shows that linear classifiers alone give linear boundaries that are suitable for the classification problem for the datasets \textit{Moons} and \textit{Linearly separable}. The MAS approach on these two datasets reproduces this linear behavior in the boundaries.

On the other hand, with the dataset \textit{Circles}, the linear models alone are unable to separate the two enclosed point clouds, contrary to the MAS approach.

\section{Discussion} \label{discussion}

Using a MAS approach, the initial classification problem is solved locally at the Context agent level. Thus, even if the agents have linear internal classifiers, they have positioned and sized each other in such a way as to locally approximate a non-linear boundary thanks to the various cooperation mechanisms presented in the section \ref{smapy}.

However, the Context agents may have over-specialized locally by observing homogeneous groups of individuals (in the sense of the class). The existence of the point exclusion mechanism, although often selected by cross validation, tends to reinforce this over-specialization of agents by excluding new class points from their activation zones.

Nevertheless, the ideal behavior sought for \textit{smapy} is to build Context agents that cover homogeneous areas of the explored input variable space, notably for reasons of explicability. There is therefore a trade-off between the geometric interpretability of the layout of the Context agents and the generalization of the system to dynamic problems in which new classes may appear.

\section{Conclusion} \label{conclusion}

Our contribution lies at  the intersection of the constructivist SACL pattern and ensemble learning methods. Our contribution provides each Context agent with an internal supervised classification model, as well as rules for cooperation with other agents. By choosing linear models for the Context agents, we show that it is possible to simplify a non-linear classification problem by transforming it into a local cooperation problem within a context learning MAS. Our experimental methodology allows us to observe a significant improvement of the classification accuracy on non linearly separable datasets.

The next step is the use of \textit{smapy} for dynamic real-world problems such as smart city management with the guarantee of an interpretable prediction compared to other state-of-the-art algorithms.

Our main research direction on \textit{smapy} is the possibility to use different algorithms in the internal models of the Context agents. The idea is to exploit the strengths and weaknesses of different known algorithms to optimize prediction quality at specific locations in the space where certain models perform best.

Finally, improvements in the operating rules of \textit{smapy} are needed to avoid over-specialization of the Context agents while maintaining the explicability and stability of the system. For this purpose, the "severity" of the NCS correction mechanisms can evolve according to the convergence of the agents' layout towards a supposedly optimal layout.

\subsubsection{Acknowledgements}

We thank the National Association for Research and Technology (ANRT) for the CIFRE funding of the thesis project in partnership with the Institut de Recherche en Informatique de Toulouse (IRIT) and the company Citec Ingénieurs Conseil. We also thank all the reviewers for their help and advice.

%
%
%
\bibliographystyle{splncs04}
\bibliography{references}

\end{document}